# Remarks on Sedov-type Solution of Isotropic Turbulence

Zheng   Ran

(Shanghai Institute of Applied Mathematics and Mechanics,
Shanghai University, Shanghai 200072, P.R.China)
e_mial: zran@staff.shu.edu.cn

**Abstract**

The assumption of similarity and self-preservation, which permits an analytical determination of the energy decay in isotropic turbulence, has played an important role in the development of turbulence theory for more than half a century. Sedov (1944), who first found an ingenious way to obtain two equations from one. Nonethless, it appears that this problem has never been reinvestigated in depth subsequent to this earlier work. In the present paper, such an analysis is carried out, yielding a much more complete picture of self-preservation isotropic turbulence. Based on these exact solutions, some physically significant consequences of recent advances in the theory of self-preserved homogenous statistical solution of the Navier-Stokes equations are presented. New results could be obtained for the analysis on turbulence features, such as the scaling behavior, the spectrum, and also the large scale dynamics. The general energy spectra and their behavior in different wave number range are investigated.

# 1. Introduction

Homogeneous isotropic turbulence is a kind of idealization for real turbulent motion, under the assumption that the motion is governed by a statistical law invariant for arbitrary translation (homogeneity), rotation or reflection (isotropy) of the coordinate system. This idealization was first introduced by Taylor (1935) and used to reduce the formidable complexity of statistical expression of turbulence and thus made the subject feasible for theoretical treatment. Up to now, a large amount of theoretical work has been devoted to this rather restricted kind of turbulence. However, turbulence observed either in nature or in laboratory has much more complicated structure. Although remarkable progress has been achieved so far in discovering various characteristics of turbulence, our understanding of the fundamental mechanism of turbulence is still partial and unsatisfactory.

The assumption of similarity and self-preservation, which permits an analytical determination of the energy decay in isotropic turbulence, has played an important role in the development of turbulence theory for more than half a century. In the traditional approach to search for similarity solutions for turbulence, the existence of a single length and velocity scale has been assumed, and then the conditions for the appearance of such solutions have been examined. Detailed research on the solutions of the Karman-Howarth equation was conducted by Sedov (1944), who showed that one could use the separability constraint to obtain the analytical solution of the Karman-Howarth equation. Sedov's solution could be expressed in terms of the confluent hypergeometric function. From the development of turbulence theory, we know that the research on decaying homogeneous isotropic turbulence is one of the most important and extensively explored topics. Despite all the efforts, a general theory describing the decay of turbulence based on the first principles has not yet been developed (Skrbek and Stalp, 2000). It seems that the theory of self-preservation in homogeneous turbulence has lots of interesting features which have not yet been fully understood and are worth of further study (see Speziale and Bernard,1992, p.665).

The first part of this work will devote to the analysis on the statistical theory of isotropic turbulence based on the Sedov-type solution. A unified investigation of isotropic turbulence is offered, based on the exact solutions of the Karman-Howarth equation. We will point out that new complete solution set may exist if we adopt the Sedov method (1944). Hence, some new results could be obtained for revealing the features of turbulence, such as the scaling behavior, energy spectra, and large-scale dynamics, and so on.

This paper focus on the two important issues:
[1] How the present work generalizes Sedov's?
[2] Dynamics in spectral space.

It should be pointed out that in the following sections, we will give some remarks on the subject respectively. More comprehensive accounts have been provided by Batchelor (1953), Hinze (1975), Monin and Yaglom (1975), Lesieur (1990), Frisch (1995), Pope (2000), and Davidson (2004), especially by the latest one.

# 2. Sedov-type solution of the Karman-Howarth equation

We will consider isotropic turbulence governed by the incompressible Navier-Stokes equations. The two-point double and triple longitudinal velocity correlation, denoted by $f(r,t)$ and $h(r,t)$ respectively, are defined in a standard way. For isotropic turbulence, they satisfy the Karman-Howarth equation

$$\frac{\partial}{\partial t}(bf) + b^{\frac{3}{2}}\left(\frac{\partial h}{\partial r} + \frac{4h}{r}\right) = 2nb\left(\frac{\partial^2 f}{\partial r^2} + \frac{4}{r}\frac{\partial f}{\partial r}\right) \tag{2.1}$$

where $(r,t)$ is the spatial and time coordinates, $n$ is the kinematic viscosity, and $b = <u^2>$ denotes the turbulence intensity.

However, these equations are not sufficient for the description of the time evolution of isotropic turbulence. Many workers have therefore tried to augment KH equation by special hypotheses containing additional information about the time variation of the correlation and spectral functions. Self-preservation hypotheses is such thing-which impose certain restrictions on the general character of the variation of the correlation and spectral functions with time. Despite the fact that isotropic turbulence constitutes the simplest type of turbulent flow, it is still not possible to render the problem analytically traceable without introducing the two point double and triple longitudinal velocity correlations to admit self-similarity solution with respect to a single length-scale, which has served as a useful hypothesis since its by von Karman and Howarth (1938). Let us suppose that the functions $f(r,t)$ and $h(r,t)$ preserve the same form as time increases with only the scale varying. Such functions will be termed as "self-preserving". Following von Karman and Howarth, we introduce the new variables

$$x = \frac{r}{l(t)} \tag{2.2}$$

where $l = l(t)$ is a uniquely specified similarity length scale. For an isotropic turbulence to be self-preserving in the sense of von Karman and Howarth (1938) and Batchelor (1948), we must have

$$f(r,t) = f(x) \tag{2.3}$$

$$h(r,t) = h(x) \tag{2.4}$$

For complete self-preservation, all scales of the turbulence, namely, those in the full range of $0 \le r \le \infty$, must decay according to (2.3) and (2.4). We will focus our attention on complete self-preserving solutions in the analysis below. For complete self-preserving isotropic turbulence, the Karman-Howarth equation takes the form

$$\frac{dh}{dx} + \frac{4}{x}h = -\frac{l}{2b^{\frac{3}{2}}}\frac{db}{dt}f + \frac{1}{2\sqrt{b}}\frac{dl}{dt}x\frac{df}{dx} + \frac{n}{\sqrt{b}l}\left(\frac{d^2 f}{dx^2} + \frac{4}{x}\frac{df}{dx}\right) \tag{2.5}$$

Sedov (1941) found an ingenious way to obtain two equations from one. The main results could be listed as following:

The two point double velocity correlation could be written as

$$\frac{d^2 f}{dx^2} + \left(\frac{4}{x} + \frac{a_1}{2}x\right)\frac{df}{dx} + \frac{a_2}{2}f = 0 \qquad (2.6)$$

subject to boundary conditions $f(0)=1$, $f(\infty)=0$, and $a_1$ and $a_2$ are constant coefficients.

The corresponding scaling (length and energy) equations read:

$$\frac{1}{\sqrt{b}}\frac{dl}{dt} = a_1\left(\frac{n}{\sqrt{bl}}\right) + p \qquad (2.7a)$$

$$-\frac{l}{b^{\frac{3}{2}}}\frac{db}{dt} = a_2\left(\frac{n}{\sqrt{bl}}\right) \qquad (2.7b)$$

where $p$ is constant of integration. These are the scaling equations of isotropic turbulence.

Equations (2.7a) and (2.7b) can be integrated to give the relations

$$\frac{1}{l} = \frac{2p}{n(a_2 - 2a_1)}\sqrt{b} + cb^{\frac{a_1}{a_2}} \qquad (2.8a)$$

for the case of $a_2 - 2a_1 \neq 0$

and

$$\frac{1}{l} = \frac{p}{n}\sqrt{b}(\ln b - c_1) \qquad (2.8b)$$

for the case of $a_2 - 2a_1 = 0$, where $c, c_1$ are the integral constants.

Sedov put $a_1 = \frac{1}{2}$; then (2.6) takes the form

$$\frac{d^2 f}{dx^2} + \left(\frac{4}{x} + \frac{1}{4}x\right)\frac{df}{dx} + \frac{a_2}{2}f = 0 \qquad (2.9)$$

Putting $a_2 = 10a > 0$ and $\frac{r^2}{l^2} = x^2 = V$, he obtained the same solution for $f(V)$ as in the small fluctuations case, namely,

$$f(V) = {}_1F_1\left(10a, \frac{5}{2}, -\frac{V}{8}\right) \qquad (2.10)$$

where ${}_1F_1(a, g, z)$ is the confluent hypergeometric function. Meanwhile, Sedov obtained

$$h = \frac{2p}{4a - 1}[f'(x) + axf(x)] \qquad (2.11)$$

Sedov also gave the solution to determine $l$ and $b$ as functions of time.

## 3. New Scaling Evolution Equations and Their Exact Solutions

In this section we deal with the scaling equations to show that how the present work generalizes Sedov's. The main part of this section is divided into four broad subsections:
1. How to derivate the new scaling equations?
2. Symmetries and power-law of the scaling equation
3. The analytical solution of the new scaling equations
4. The corresponding induced turbulence decay laws

### 3.1 Derivation of the new scaling equation

For complete self-preserving isotropic turbulence, the corresponding scaling equation takes the form of Eq.(2.7). Unlike what Sedov (1944) has done, we can only obtain a closed equation for the length scale $l(t)$. These equations can be recast into

$$\frac{dl}{dt} = a_1\left(\frac{n}{l}\right) + p\sqrt{b} \quad (3.1a)$$

$$\frac{db}{dt} = -a_2\left(\frac{nb}{l^2}\right) \quad (3.1b)$$

By using eq.(3.1a), we have

$$\frac{d^2l}{dt^2} = -\frac{a_1 n}{l^2} \cdot \frac{dl}{dt} + \frac{p}{2} \cdot \frac{1}{\sqrt{b}} \cdot \frac{db}{dt} \quad (3.2)$$

In any event, substituting (3.1b) into our equation (3.2) yields

$$\frac{d^2l}{dt^2} = -\frac{a_1 n}{l^2} \cdot \frac{dl}{dt} + \frac{p\sqrt{b}}{2} \cdot \frac{1}{b} \cdot \frac{db}{dt}$$
$$= -\frac{a_1 n}{l^2} \cdot \frac{dl}{dt} - \frac{a_2 n}{2l^2} \cdot \left\{\frac{dl}{dt} - \frac{a_1 n}{l}\right\} \quad (3.3)$$

After a little work, we have

$$\frac{d^2l}{dt^2} + \frac{(2a_1 + a_2)n}{2l^2}\frac{dl}{dt} - \frac{a_1 a_2 n^2}{2l^3} = 0 \quad (3.4)$$

If we know the length scale, the substituting equation (3.1b) will give the turbulence energy decay law. So, the departure point is how to deal with the equation (3.4).

### 3.2 Symmetries of the scaling equation
Let

$$P(l) = \frac{dl}{dt} \quad (3.5)$$

Equ.(3.4) can recast into

$$PP'_l = -\frac{(2a_1 + a_2)n}{2l^2} \cdot P + \frac{a_1 a_2 n^2}{2l^3} \tag{3.6}$$

$$P'_l = -\frac{(2a_1 + a_2)n}{2l^2} \cdot + \frac{a_1 a_2 n^2}{2l^3} \cdot P^{-1} \tag{3.7}$$

Symmetries and first integrals are two fundamental structures of ordinary differential equations (ODEs). Geometrically, it is natural to view an nth-order ODE as a surface in the $(n+2)$-dimensional space whose coordinates are given by the independent variable, the dependent variable and its derivates to order $n$, so that the solutions of the ODE are particular curves lying on this surface. From this point of view, symmetry represents a motion that moves each solution curve into solution curves; a first integral represent a quantity that is conserved along each solution curve. More precisely, symmetry is a one-parameter group of local transformations, acting on the coordinates involving the independent variable, the dependent variable and its derivatives to order $n-1$, that is constant on each solution. In this section, we show how to find admitted symmetries and first integrals of the turbulence scaling equation.

Based on the Lie symmetry analysis, equation (3.7) only admits a one-parameter Lie group of point transformation with infinitesimal (see Appendix 1)

$$V = \frac{l}{\bar{a}} \frac{\partial}{\partial l} + \frac{P}{\bar{b}} \frac{\partial}{\partial P} \tag{3.10}$$

where $\bar{a}, \bar{b}$ are arbitrary constants satisfies

$$\frac{1}{\bar{a}} + \frac{1}{\bar{b}} = 0. \tag{3.11}$$

The corresponding characteristic equation is given by

$$\frac{\bar{a}}{l} dl = \frac{\bar{b}}{P} dP \tag{3.12}$$

The invariant

$$I_1 = \frac{P^{\bar{b}}}{l^{\bar{a}}} \tag{3.13}$$

This gives

$$P = \frac{dl}{dt} = [I_1]^{\frac{1}{\bar{b}}} \cdot l^{-1} \tag{3.14}$$

The solution in term of time could be

$$l^2 = 2[I_1]^{\frac{1}{\bar{b}}} \cdot (t + t_0) \tag{3.15}$$

It is well known that if an ODE admits a Lie group of transformations, then one can construct interesting special classes of solutions (invariant solutions) that correspond to invariant curves of the admitted by Lie group of transformations. For a first-order ODE, such invariant solutions can be determined algebraically. It is believed that this might be show some light on the nature of the power-law in scaling equation.

## 3.3 The solution of the first integrals

Based on the result of section 3.2, we are interested in the special solutions in the following form:

$$l(t) = l_0 (t + t_0)^{\frac{1}{2}} \tag{3.16}$$

$$b(t) = b_0 (t + t_0)^s \tag{3.17}$$

where $l_0, b_0, s$ are parameters to be chosen. Special attention would be paid to the derivation of $l_0$. From equation (3.16), we have the time derivatives

$$\frac{dl}{dt} = \frac{l_0}{2} \cdot (t + t_0)^{-\frac{1}{2}} \tag{3.18}$$

$$\frac{d^2 l}{dt^2} = -\frac{l_0}{4} \cdot (t + t_0)^{-\frac{3}{2}} \tag{3.19}$$

The substitution eq. (3.18)+eq.(3.19) into eq.(3.4) yields

$$-\frac{l_0}{4} \cdot (t+t_0)^{-\frac{3}{2}} + \frac{(2a_1 + a_2)n}{2l_0^2} \cdot \frac{l_0}{2} \cdot (t+t_0)^{-\frac{3}{2}} - \frac{a_1 a_2 n^2}{2l_0^3} \cdot (t+t_0)^{-\frac{3}{2}} = 0 \tag{3.20}$$

For general case

$$t + t_0 > 0 \tag{3.21}$$

So, we have

$$-\frac{l_0}{4} + \frac{(2a_1 + a_2)n}{2l_0^2} \cdot \frac{l_0}{2} - \frac{a_1 a_2 n^2}{2l_0^3} = 0 \tag{3.22}$$

This equation can be recast into

$$l_0^4 - (2a_1 + a_2)n l_0^2 + 2a_1 a_2 n^2 = 0 \tag{3.23}$$

This leads

$$l_0^2 = \frac{n}{2} \left\{ (2a_1 + a_2) \pm \sqrt{(2a_1 - a_2)^2} \right\} \tag{3.24}$$

Here, we would like to introduce some notations for the convenience.
If

$$l_0^2 = \frac{n}{2} \left\{ (2a_1 + a_2) + \sqrt{(2a_1 - a_2)^2} \right\} \tag{3.25}$$

we denote its by P_mode.
If

$$l_0^2 = \frac{n}{2} \left\{ (2a_1 + a_2) - \sqrt{(2a_1 - a_2)^2} \right\} \tag{3.26}$$

we denote its by N_mode.

But the final results will depend on the values of $s$. The definition of $s$ is

$$s \equiv \frac{a_2}{2a_1} \tag{3.27}$$

The details of the solutions are listed in Table.1 and 2.

## Table.1 Values of $l_0^2$

|  | **P_mode** | **N_mode** |
|---|---|---|
| $0 < s \leq 1$ | $2a_1 n$ | $a_2 n$ |
| $s > 1$ | $a_2 n$ | $2a_1 n$ |

## Table.2 Values of $s$

|  | **P_mode** | **N_mode** |
|---|---|---|
| $0 < s \leq 1$ | $-s$ | $-1$ |
| $s > 1$ | $-1$ | $-s$ |

Grid turbulence was first extensively studied by Batchelor and Townsend (1948). The turbulence, was observed to have a power-law decay in energy. A detailed study of self-similar solutions of the Karman-Howarth equation and their stable equilibria can be found in Speziale and Benard (1992). They conclude that completely self-similar solutions always lead to a decay exponent of one, unless the Reynolds number is zero, a state that can be reached only as time goes to infinity. However, as demonstrated in present paper, asymptotic decay laws other than one must be due to the fact that the new scaling equation has two different kinds of invariant solutions.

**3.4 The alternate analytical solution**

In this subsection, an alternate route of obtaining the exact solution for the new scaling equation will be taken. It is important to note that equation (3.4) is a second class of nonlinear Lienard type equation and one can obtain another exact solution by using the standard method. (Appendix 2.)

The start point is the equation (3.7), furthermore, by using the methods presented in the Appendix 2, we have the expressions of the corresponding functions as following:

$$F(l) = -\frac{(2a_1 + a_2)n}{2l^2} \tag{3.28}$$

$$G(l) = \frac{a_1 a_2 n^2}{2l^3} \tag{3.29}$$

Introduce the transformation of variables

$$Z = \int F(l)dl \tag{3.30}$$

Hence

$$Z = \frac{(2a_1 + a_2)n}{2} \cdot \frac{1}{l} + Z_0 \tag{3.31}$$

where $Z_0$ is integral constant.

Hence,

$$l = \frac{(2a_1 + a_2)n}{2} \cdot \frac{1}{Z - Z_0} \tag{3.32}$$

So, we obtain the canonical equation

$$PP'_Z = P + \Phi(Z) \tag{3.33}$$

where

$$\begin{aligned}\Phi(Z) &= \frac{G(Z)}{F(Z)} = -\frac{a_1 a_2}{(2a_1 + a_2)} \cdot \frac{n}{l} \\ &= -\frac{2a_1 a_2}{(2a_1 + a_2)^2}(Z - Z_0) \\ &= -\frac{2a_1 a_2}{(2a_1)^2\left(1 + \dfrac{a_2}{2a_1}\right)^2} \cdot (Z - Z_0) \\ &= -\frac{s}{(1+s)^2}(Z - Z_0)\end{aligned} \tag{3.34}$$

At last we obtain a new canonical equation for the scaling evolution as

$$PP'_Z = P - \frac{s}{(1+s)^2}(Z - Z_0) \tag{3.35}$$

Let

$$Y = Z - Z_0 \tag{3.36}$$

We have

$$PP'_Y = P - \frac{s}{(1+s)^2}Y \tag{3.37}$$

This is the fact as following

$$P'_z = \frac{dP}{dY} \cdot \frac{dY}{dZ} = \frac{dP}{dY} \tag{3.38}$$

By comparison the method presented by the Appendix 2, and detail calculation of this equation, we have

$$l = \sqrt{(2a_1 + a_2)w^* n} \cdot \sqrt{t + t_0}. \qquad (3.39)$$

Here $w^*$ is one parameter which comes from the mean value theorems. Based on this conclusion, the turbulent energy easily obtained as

$$b = b_0 (t + t_0)^{-\frac{s}{(1+s)w^*}} \qquad (3.40)$$

**3.5 The turbulence decay law**

So far we have focus on the different exact solutions of the length scaling equation. We now switch to the turbulent energy. Combine the above two different exact solutions, we have the new decay law, written as:

$$\bar{b} = \frac{b_0}{2}\{b_1 + b_2\} \qquad (3.41)$$

where $b_1$ denotes the solution independent on the initial condition, and $b_2$ denotes the solution dependents on the initial condition. They are listed in Table.3 and equation (3.41) respectively. This indicates that we must to use two exponents to describe their features. Here, we introduce $(-s, -q)$ to describe this set of exponents. The behavior dependence on the parameter $s$ could be given as following

**Table.3 Values of the exponents of turbulent energy**

|  | **P_mode** | **N_mode** |
|---|---|---|
| $0 < s \leq 1$ | $(-s, -q)$ | $(-1, -q)$ |
| $s > 1$ | $(-1, -q)$ | $(-s, -q)$ |

For the P_mode:
$s > 1$:

$$\bar{b} = \frac{b_0}{2} \cdot \frac{1}{(t+t_0)^s} + \frac{b_0}{2} \cdot \frac{1}{(t+t_0)^q} \qquad (3.42)$$

$0 < s \leq 1$:

$$\bar{b} = \frac{b_0}{2} \cdot \frac{1}{(t+t_0)^1} + \frac{b_0}{2} \cdot \frac{1}{(t+t_0)^q} \qquad (3.43)$$

For the N_mode:
$0 < s \leq 1$:

$$\bar{b} = \frac{b_0}{2} \cdot \frac{1}{(t+t_0)^s} + \frac{b_0}{2} \cdot \frac{1}{(t+t_0)^q} \qquad (3.44)$$

$s > 1$:

$$\bar{b} = \frac{b_0}{2} \cdot \frac{1}{(t+t_0)^1} + \frac{b_0}{2} \cdot \frac{1}{(t+t_0)^q} \qquad (3.45)$$

where

$$q = \frac{s}{(1+s)w^*}$$

From above analysis, we can give the most suitable expression of the turbulent energy,

$$\langle b \rangle \equiv \frac{b_0}{3} \cdot \frac{1}{(t+t_0)^1} + \frac{b_0}{3} \cdot \frac{1}{(t+t_0)^s} + \frac{b_0}{3} \cdot \frac{1}{(t+t_0)^q} \qquad (3.46)$$

reflecting the dependence of distribution of parameter $s$, where $\langle b \rangle$ represents the average value of the turbulent energy.

Meanwhile, we also notice that:
[1] In 1958, Deissiler [1] used a three-point correlation equation to obtain a relation for the triple correlations applicable at times before the final period. In this case the equation is made by neglecting the quadruple correlation. By integrating the energy spectrum over all wave numbers (or eddy sizes), the following energy decay law was obtained

$$\bar{b} = A \cdot \frac{1}{(t+t_0)^{\frac{5}{2}}} + B \cdot \frac{1}{(t+t_0)^7} \qquad (3.47)$$

where $A, B$ and $t_0$ are constants determined by the initial conditions.

[2] Eq.(3.46) is remindful of analytic solutions to the one point equations that Ristorcelli (2003) and someone else did. They showed that there was no single power law for the decay except perhaps asymptotically.

To summarize, the main conclusion of this section appears to be the decay law given in eq. (3.46), due to the three different type solutions of the length scaling equation. Eq. (3.46) finds what looks like an infinity of possibilities. It is very unexpected because complete self-similarity is conventionally said to a decay exponent of one. This is what the present work has done beyond Sedov's work, noting that complete self-similarity permits a richer class of solutions than previously believed.

---

[1] Deissiler, R.G., On the decay of homogeneous turbulence before the final period. Phys. Fluids., vol.1, no.2, pp.111-121, 1958.

# 4. Dynamics in spectral space

For isotropic turbulence, the Karman-Howarth equation, which stems from the Navier-Stokes equations, fully describes the dynamics of the two-point velocity correlation. It does not, however, provide a very clear picture of the processes involved in the energy cascade. Some further insights can be gained by examining the Navier-Stokes equations in the wave-numbers space. In this section, we will examine the energy spectrum of isotropic turbulence based on the exact solution. Several functional forms for the energy spectrum in small wave range, inertial-range have been proposed based on the asymptotic analysis.

## 4.1 Spectra based on the Sedov-type solution

In the following analysis, we introduce two alternative parameters denoted by $a_1, s$, while the Sedov-type solution could be rewritten as

$$f(x) = e^{-\frac{a_1}{4}x^2} {}_1F_1\left(\frac{5}{2} - s, \frac{5}{2}, \frac{a_1}{4}x^2\right) \tag{4.1}$$

One-dimensional energy spectra could be deduced directly from this solution.

A turbulent flow varies randomly in all three space direction and in time. Experimental measurements, say of velocity, may be made along a straight line at a fixed time, at a fixed position as function of time, or following a moving fluid point as function of time. A measurement of this kind generates a random function of position or time. If the function is stationary or homogeneous, an autocorrelation can be formed and a spectrum can be computed. If the autocorrelation is a function of a time interval, the transform variable is wave number. Spectra obtained in this way are called one-dimensional spectra because the measurements production them were taken in one dimension. The one-dimensional spectra that are most often measured are the one-dimensional Fourier transforms of a longitudinal or transverse correlation. There is no uniformity of notation for one-dimensional energy spectra. Batchelor (1953), Hinze (1975), Tennekes and Lumley (1972), and Monin and Yaglom (1975) all use different symbols. We shall adopt the same convention as Tennekes and Lumley for the one-dimensional spectra of $u^2 f$ and $u^2 g$, that is $F_{11}$ and $F_{22}$. (Davidson, 2004)

They appear particularly in experimental papers as the quantities most commonly measured in experiment. They are

$$F_{11}(k,t) = \frac{1}{p} \int_0^\infty u^2 f(r,t) \cos(kr) dr \tag{4.2}$$

$$F_{22}(k,t) = \frac{1}{p} \int_0^\infty u^2 g(r,t) \cos(kr) dr \tag{4.3}$$

with inverse transform,

$$u^2 f(r,t) = 2\int_0^\infty F_{11}(k,t)\cos(kr)dk \tag{4.4}$$

$$u^2 g(r,t) = 2\int_0^\infty F_{22}(k,t)\cos(kr)dk \tag{4.5}$$

Here $g$ and $f$ are the usual transverse and longitudinal correlation functions. Of course, $F_{11}(k,t)$ $F_{22}(k,t)$ are simply the one-dimensional energy spectra of $u_x(x,0,0)$ and $u_y(x,0,0)$.

By using the integral formula:

$$\int_0^\infty {}_1F_1(a;c;-t^2)\cos(2zt)dt = \sqrt{\frac{p}{2}}\frac{\Gamma(c)}{\Gamma(a)}z^{2a-1}e^{-z^2}U\left(c-\frac{1}{2},a+\frac{1}{2},z^2\right) \tag{4.6}$$

where ${}_1F_1(a,c,z)$ is Kummer's hypergemometric function, and $U(a,c,z)$ is another function closely related to Kummer's function defined by

$$U(a,c,z) = \frac{\Gamma(1-c)}{\Gamma(1+a-c)}{}_1F_1(a,c,z) + \frac{\Gamma(c-1)}{\Gamma(a)}z^{1-c}{}_1F_1(1+a-c,2-c,z) \tag{4.7}$$

where $\Gamma(z)$ is the usual Gamma function.

For the Sedov-type solution, according to formula (4.5), for isotropic turbulence, we have

$$F_{11}(k,t) = \frac{2u^2}{p}\cdot\frac{l}{\sqrt{a_1}}\cdot\sqrt{\frac{p}{2}}\frac{\Gamma\left(\frac{5}{2}\right)}{\Gamma(s)}\cdot\left(\frac{l}{\sqrt{a_1}}k\right)^{2s-1}e^{-\frac{l^2}{a_1}k^2}U\left(2,s+\frac{1}{2},\frac{l^2}{a_1}k^2\right) \tag{4.8}$$

$$E(k,t) = \sqrt{\frac{8}{pa_1}}\cdot\left(\frac{2}{a_1}\right)^2\cdot\frac{\Gamma\left(\frac{5}{2}\right)}{\Gamma(s)}\cdot(bl)\cdot(kl)^4\cdot e^{-\frac{l^2}{a_1}k^2}\cdot U\left(\frac{5}{2}-s,\frac{7}{2}-s,\frac{l^2}{a_1}k^2\right) \tag{4.9}$$

This is the exact result of the spectra based on the Sedov-type solution. In further consideration, we want to know it is probably fair to say: Is this a satisfactory closure scheme which encompasses both the large and the small scales?

**4.2 Large-scales dynamics of isotropic turbulence**

A third of a century has passed since Saffman's (1967) paper, yet there is still controversy over whether turbulence is of the Batchelor spectra, or Saffman spectra. It seems that both are, in principle, realizable and that either may be generated in computer simulations. The real question, of course, is which does nature prefer? It turns out that the initial conditions are crucial in this respect (Davidson, 2004). By using the asymptotic analysis of equ.(4.9), it is easy to obtain the large scales dynamics of isotropic turbulence.

By using the definition of $U(a,c,z)$, we also have[2]

$$U\left(\frac{5}{2}-s, \frac{7}{2}-s, z\right) = U\left(\frac{5}{2}-s, \frac{5}{2}-s+1, z\right) \tag{4.10}$$

We also note that

$$(a+z)U(a,c,z) + a(c-a-1)U(a+1,c,z) - zU(a,c+1,z) = 0 \tag{4.11}$$

$$U(a,c+1,z) = \left(1+\frac{a}{z}\right)U(a,c,z) + a(c-a-1)\cdot\frac{1}{z}\cdot U(a+1,c,z) \tag{4.12}$$

The substituting yields

$$E(k,t) = \sqrt{\frac{8}{pa_1}} \cdot \left(\frac{2}{a_1}\right)^2 \cdot \frac{\Gamma\left(\frac{5}{2}\right)}{\Gamma(s)} \cdot (bl) \cdot (kl)^4 \cdot e^{-\frac{l^2}{a_1}k^2} \times$$

$$\left\{\left[1 + \frac{\frac{5}{2}-s}{\frac{(kl)^2}{a_1}}\right] U\left(\frac{5}{2}-s, \frac{5}{2}-s, \frac{l^2}{a_1}k^2\right) - \frac{\frac{5}{2}-s}{\frac{(kl)^2}{a_1}} \cdot U\left(\frac{7}{2}-s, \frac{5}{2}-s, \frac{l^2}{a_1}k^2\right)\right\} \tag{4.13}$$

Based on the above formula, it is easy to handle the asymptotic analysis on the behavior of the energy spectra at the low wave numbers.

So, if $s \neq \frac{5}{2}$, the simplest asymptotic function forms are listed as follows:

[1]For $s > \frac{3}{2}, s \neq \frac{5}{2}, k \to 0$:

$$U\left(\frac{5}{2}-s, \frac{5}{2}-s, \frac{l^2}{a_1}k^2\right) = \frac{\Gamma\left(s-\frac{3}{2}\right)}{\Gamma(1)} \tag{4.14}$$

$$U\left(\frac{7}{2}-s, \frac{5}{2}-s, \frac{l^2}{a_1}k^2\right) = \frac{\Gamma\left(s-\frac{3}{2}\right)}{\Gamma(2)} \tag{4.15}$$

The substitutions of equ.(4.14)+(4.15) into (4.13) lead

---

[2] This mathematical manipulation lies in the fact that the expression equ.(4.9) is not well defined since it has a pole $\Gamma(0)$, according to the definition (4.7), one can attempt to define new expression of the function by using recurrence relations to elimination this term.

$$E(k,t) = \sqrt{\frac{8}{pa_1}} \cdot \left(\frac{2}{a_1}\right)^2 \cdot \frac{\Gamma\left(\frac{5}{2}\right)}{\Gamma(s)} \cdot (bl) \cdot (kl)^4 \cdot e^{-\frac{l^2}{a_1}k^2} \times$$

$$\left\{\left[1 + \frac{\frac{5}{2} - s}{\frac{(kl)^2}{a_1}}\right] \cdot \Gamma\left(s - \frac{3}{2}\right) - \frac{\frac{5}{2} - s}{\frac{(kl)^2}{a_1}} \cdot \Gamma\left(s - \frac{3}{2}\right)\right\} \quad (4.16)$$

$$= \sqrt{\frac{8}{pa_1}} \cdot \left(\frac{2}{a_1}\right)^2 \cdot \frac{\Gamma\left(s - \frac{3}{2}\right)}{\Gamma(s)} \cdot \Gamma\left(\frac{5}{2}\right) \cdot (bl^5) \cdot k^4$$

This recovers the former Batchelor-spectra.

[2]**For** $\frac{1}{2} < s < \frac{3}{2}$, $k \to 0$:

$$U\left(\frac{5}{2} - s, \frac{5}{2} - s, \frac{l^2}{a_1}k^2\right) = \frac{\Gamma\left(\frac{3}{2} - s\right)}{\Gamma\left(\frac{5}{2} - s\right)} \cdot \left(\frac{l^2}{a_1}k^2\right)^{s - \frac{3}{2}} \quad (4.17)$$

$$U\left(\frac{7}{2} - s, \frac{5}{2} - s, \frac{l^2}{a_1}k^2\right) = \frac{\Gamma\left(\frac{5}{2} - s\right)}{\Gamma\left(\frac{7}{2} - s\right)} \cdot \left(\frac{l^2}{a_1}k^2\right)^{s - \frac{3}{2}} \quad (4.18)$$

Hence,

$$E(k,t) = \sqrt{\frac{8}{pa_1}} \cdot \left(\frac{2}{a_1}\right)^2 \cdot \frac{\Gamma\left(\frac{5}{2}\right)}{\Gamma(s)} \cdot (bl) \cdot (kl)^4 \cdot e^{-\frac{l^2}{a_1}k^2} \times$$

$$\left\{\left[1 + \frac{\frac{5}{2} - s}{\frac{(kl)^2}{a_1}}\right] \cdot \frac{1}{\frac{3}{2} - s} - \frac{\frac{5}{2} - s}{\frac{(kl)^2}{a_1}} \cdot 1\right\} \cdot (kl)^{2s-3} \quad (4.19)$$

$$= \sqrt{\frac{8}{pa_1}} \cdot \left(\frac{2}{a_1}\right)^2 \cdot \frac{\Gamma\left(s - \frac{3}{2}\right)}{\Gamma(s)} \cdot \Gamma\left(\frac{5}{2}\right) \cdot a_1 \cdot \left(\frac{5}{2} - s\right) \cdot \frac{2s - 1}{3 - 2s} \cdot (bl^{2s}) \cdot k^{2s-1}$$

[3]For $s = \frac{3}{2}$ [3], this recovers the former Saffman-spectra. To obtain the right expression, we

---

[3] $U(a,1,z) = -\frac{1}{\Gamma(a)}\{\log z + \Psi(a) - 2g\}$, $g$ =Euler's constant,

$\Psi(x)$ is psi function.

must consider the following formula:

$$U\left(1,1,\frac{l^2}{a_1}k^2\right) = -\frac{1}{\Gamma(1)}\left\{\log\left[\frac{l^2}{a_1}k^2\right] + \Psi(1) - 2g\right\} \quad (4.20)$$

$$U\left(2,1,\frac{l^2}{a_1}k^2\right) = -\frac{1}{\Gamma(2)}\left\{\log\left[\frac{l^2}{a_1}k^2\right] + \Psi(2) - 2g\right\} \quad (4.21)$$

Hence,

$$E(k,t) = \sqrt{\frac{8}{pa_1}} \cdot \left[\frac{2}{a_1}\right]^2 \cdot \frac{2}{3} \cdot a_1 \cdot \{\Psi(2) - \Psi(1)\} \cdot (bl^3) \cdot k^2 \quad \textbf{(4.22)}$$

The main results retrieved from the present analysis are as follows:

[1] The parameter $s$ serves as the unique classification index of the asymptotic behavior of turbulent spectra in the low wave number range.

[2] The analysis previously conducted relies on the Sedov-type solution, and demonstrates that a self-similar state whose low wave number kinetic energy spectrum is

$$E(k,t) \propto C_s(t) \cdot I \cdot k^n$$

The corresponding time evolution of the turbulent kinetic energy spectrum is displayed in the table 4.1.

**Table.4.1 Time evolution exponents in self-similar decay of isotropic turbulence deduced from present analysis**

|  | spectra | Saffman spectra | Batchelor spectra |
|---|---|---|---|
| $s$ | $s < \frac{3}{2}$ | $s = \frac{3}{2}$ | $s > \frac{3}{2}$ |
| $C_s$ | $C_-$ | $C_0$ | $C_+$ |
| $I$ | $bl^{2s}$ | $bl^3$ | $bl^5$ |
| $n$ | $2s - 1$ | 2 | 4 |

**Where**

$$C_- = \sqrt{\frac{8}{pa_1}} \cdot \left[\frac{2}{a_1}\right]^2 \cdot \frac{\Gamma\left(\frac{5}{2}\right)}{\Gamma(s)}$$

$$C_0 = \frac{2}{3}a_1 \cdot \sqrt{\frac{8}{pa_1}} \cdot \left[\frac{2}{a_1}\right]^2 \cdot \{\Psi(2) - \Psi(1)\}$$

$$C_+ = \sqrt{\frac{8}{pa_1}} \cdot \left[\frac{2}{a_1}\right]^2 \cdot \frac{\Gamma\left(\frac{5}{2}\right)}{\Gamma(s)}$$

### 4.3 The power-law spectra

Sixty years ago, .A.N.Kolmogorov (1941) proposed an elegant theory of the universal statistical properties of small-scale eddies in high Reynolds number turbulent flows. Kolmogorov's theory (hereafter referred to as K41) is still the basis for nearly all work on the statistical theory of turbulence. By using physically motivated and dimensional arguments, spectrum estimates of some velocity moments have been obtained. Of course those arguments had not referred explicitly to the equations governing turbulent flows; a notable example is original "derivation" of the energy spectrum of full developed homogeneous isotropic turbulence by Kolmogorov. Curiously enough, little attention has been paid to the possibility of eliciting physically meaningful result directly from the properties of the statistics governing turbulence. It is worth stressing that K41 makes no direct connection to the Navier-Stokes equations, furthermore, K41 is also not correct in detail (Sreenivassan[4], 1999).

The main task in connection with the theory of isotropic turbulence at present seems to be the prediction or explanation of this power law. This is still the main challenge for isotropic turbulence theory. This section devotes to this important issue. We will see that the power-law is one asymptotic state, derived naturally from the asymptotic expansion of the general turbulent spectra given above.

We remember that, the turbulent spectrum has been rewritten as eq. (4.13). It is natural to see what will follow if we adopt the asymptotic expansion in the case of the large argument. The details have been listed from the mathematical point of view. Furthermore, we also have known that:

For the asymptotic expansions of the function $U(a,c,z)$, as $z \to \infty$, we have

$$U\left(\frac{5}{2} - s, \frac{5}{2} - s, \frac{l^2}{a_1}k^2\right) = \sum_{m=0}^{N} A_m \left(\frac{l^2}{a_1}k^2\right)^{-m-\left(\frac{5}{2}-s\right)} \quad (4.23)$$

$$U\left(\frac{7}{2} - s, \frac{5}{2} - s, \frac{l^2}{a_1}k^2\right) = \sum_{m=0}^{N} B_m \left(\frac{l^2}{a_1}k^2\right)^{-m-\left(\frac{7}{2}-s\right)} \quad (4.24)$$

where

$$A_m = (-1)^m \cdot \frac{\left(\frac{5}{2} - s\right)_m \cdot (1)_m}{m!} \quad (4.25)$$

---

[4] K.R.Sreenivassan, Fluid turbulence. Reivew of Modern Physics, vol.71, no.2, 1999.

$$B_m = (-1)^m \cdot \frac{\left(\frac{7}{2}-s\right)_m \cdot (2)_m}{m!} \tag{4.26}$$

Performing some mechanical calculations, we have

$$E(k,t) = \sum_{m=0}^{M} \sqrt{\frac{8}{pa_1}} \cdot \left(\frac{2}{a_1}\right)^2 \cdot \frac{\Gamma\left(\frac{5}{2}\right)}{\Gamma(s)} \cdot \left(bl^{2(s-m)-2}\right) \cdot e^{-\frac{l^2}{a_1}k^2} \times \left\{ \left[\frac{\frac{5}{2}-s}{\frac{1}{a_1}}\right] \cdot A_m \cdot \left(\frac{1}{a_1}\right)^{(s-m)-\frac{5}{2}} \right\} \cdot k^{2(s-m)-3} \tag{4.27}$$

It should not be forgotten, however, that the entire discussion above is restricted to recover the power law spectra. While it is naturally to set that:

$$2(s-m) - 3 = -\frac{5}{3} \tag{4.28}$$

Hence,

$$s_m = \frac{2}{3} + m \tag{4.29}$$

where $m = 0,1,2,....,M$.

The corresponding energy spectrum takes the form,

$$E(k,t) = C_L \cdot J(t) \cdot k^{-\frac{5}{3}} \tag{4.30}$$

where[5]

$$C_L = \sum_{m=0}^{M} \sqrt{\frac{8}{pa_1}} \cdot \left(\frac{2}{a_1}\right)^2 \cdot \frac{\Gamma\left(\frac{5}{2}\right)}{\Gamma(s)} \left\{ \left[\frac{\frac{5}{2}-s}{\frac{1}{a_1}}\right] \cdot A_m \cdot \left(\frac{1}{a_1}\right)^{-\frac{11}{6}} \right\} > 0 \cdot \tag{4.31}$$

$$J(t) = bl^{-\frac{2}{3}} \tag{4.32}$$

Now it is time we learn more things from K41. For small scales much larger than the Kolmogorov scale, one recovers the inertial range expression:

$$E_K(k) = K_0 \cdot e^{\frac{2}{3}} \cdot k^{-\frac{5}{3}} \tag{4.33}$$

where $K_0$ is the Kolmogorov canstant.

It is important to note that in the present framework,

---

[5] One can easily confirm that $C_L > 0$.

$$e = e_0\left(bl^{-2}\right) \tag{4.34}$$

After a little work, one finds

$$E(k,t) = C_L \cdot \left(bl^2\right)^{\frac{1}{3}} \cdot e^{\frac{2}{3}} \cdot k^{-\frac{5}{3}} \tag{4.35}$$

This is the final expression of the power law spectrum based on the exact solution. The connection of this conclusion with Kolmogorov's theory in inertial range is an interesting question. So, the main effects are readily recovered in our analysis.

In comparison, it seems that the conclusion drawn from this section could be listed as following:

[1]It is natural to come to the K41-like power-law spectra, if we adopt the asymptotic expansion in the case of the large argument and set $s_m = \frac{2}{3} + m$; $m = 0,1,2,...,M$;

[2]It is important to note that the "constants" involved in these decay laws are to be dependent of time. This feature represents a crucial departure from K41.

## 5. Consistency of the different spectra

In this section, we focus on the consistency problem in isotropic turbulence. Based on the results above, we have K41-like theory of the small scales, theory of the large scales. They are two different asymptotic states. So, after such trip, where do we stand in our understanding of isotropic turbulence for the entire wave number range? In any event, this is a question. It is well known that: For the large scale dynamics, there has been much controversy over these various findings, and even today there is little agreements to whether grid turbulence is of the Saffman type or of the Batchelor type. Things are the same for the small scale dynamics, as witnessed by the lack of full satisfactory theories of such basic aspects as the K41.

Let us now take matters a step further. We recall that:

[1]The parameter $s$ serves as the unique classification index of the asymptotic behavior of turbulent spectra in the low wave number range.

[2]It is natural to come to the K41-like power-law spectra, if we adopt the asymptotic expansion in the case of the large argument and set $s_m = \frac{2}{3} + m$; $m = 0,1,2,...,M$.

It is not difficult to show that

$$s_0 = \frac{2}{3} \tag{5.1}$$

$$s_1 = \frac{2}{3} + 1 \tag{5.2}$$

$$s_2 = \frac{2}{3} + 2 \tag{5.3}$$

$$\ldots$$

$$s_M = \frac{2}{3} + M \tag{5.4}$$

Note that

$$s_c = \frac{3}{2} \tag{5.5}$$

We have seen that

$$s_0 < s_c < s_1 < .... < s_M \tag{5.6}$$

Let us now consider the results of the table 4.1. We learn two more things from this table.

[1] $m = 0$, $k \to 0$,

$$E(k,t) = C_-(t) \cdot \left( bl^{\frac{4}{3}} \right) \cdot k^{\frac{1}{3}} \tag{5.7}$$

[2] $m = 1, 2, ..., M$, $k \to 0$,

$$E(k,t) = C_+(t) \cdot \left( bl^5 \right) \cdot k^4 \tag{5.8}$$

It tells us some about the large scale dynamics.

So, for a while, it looked like we had a sound theory of both the large and the small scales.

## 6. Conclusion and discussion

We have revisited the old problem firstly presented by Sedov (1944), and found richer mathematical structure in this paper compared to Sedov's work. The results help us to offer a unified investigation of isotropic turbulence. This viewpoint allows us to classify the theoretical approaches to isotropic turbulence as follows:

[1]The central idea is the locality of the interaction, as proposed by Kolmogorov in describing the incompressible fluids: only vortices of spatial extensions of the same order strongly interact with each other. Kolmogorov's hypothesis leads to a lot of activities and to an immense scientific literature. Experimenters essentially confirm the validation. A rigorous theoretical substantiation of the K41 is not yet available. This is mainly due to the absence of a small parameter in the traditional theory of hydrodynamic turbulence. Vortex interaction in incompressible fluids is strong and there exists no small parameter in hydrodynamic equations. According to our physical picture of turbulence, this small parameter may be $s$, serving as unique classification index of turbulent flows. It comes from the Karman-Howarth equation, reflecting the statistical nature.

[2]Finite Reynolds number corrections could also be regarded as a natural result from the asymptotic expansions of the general turbulent spectra in different asymptotic conditions.

[3]For the present theory, the coefficient of the enstrophy destruction $G(t)$ will be a constant(but dependent on the distribution of the parameter $s$ indeed). Analytical study of present theory may be useful in understanding the nature of isotropic turbulence. All of the results presented here are exact and unified, giving deep insight of the internal structure of isotropic turbulence.

[4]The results obtained confirm the qualitatively consistency of the approximation and suggest a satisfactory quantitative agreement with experiment in the range of $R_l$ which is treated. Much more through confrontation with experiment is need to confirm this conclusion, however. Analytical study of present theory may be useful in understanding the nature of isotropic turbulence. The values of $S(t)$ shown in this paper appears to be consistent with experiment values and other turbulence theories, such as DIA, EDQNM , qualitatively.(Ran,2008,2009)

It is worth noting that: questions about Sedov's work. The main fact about the Karman-Howarth equations is that it is one equation for two unknowns. Sedov found an ingenious way to obtain two equations from one. A determinate result requires the viscous preservation in Sedov's approach. The introduction of viscosity represents a crucial departure from the conventional thinking. There are still several directions where the Sedov's work calls for substantial progress and upgrades.

Another important problem we must mention here. The ubiquity of expressions of turbulence energy spectrum containing factor $\exp(-k^2)$ in present work, suggests a linearly damped dissipation range; however, modern studies of the dissipation range beginning with Kraichnan (1959) agree that the dissipation range scales as $\exp(-k^1)$ as the result of nonlinearity acting against viscosity, confirmed by numerical studies. This is still a matter of some controversy.

**Acknowledgements:**
The author is indebted to the referees for their helpful comments on the manuscript. The work was supported by the National Natural Science Foundation of China (Grant Nos.10272018, 10572083), and also supported by Shanghai Leading    Academic Discipline Project, Project Number: Y0103.

# Appendix 1: Symmetry analysis of the scaling equation

It will be shown that if this equation is invariant under a one-parameter group of transformations, then it can be integrated by quadrature. We discuss the first order differential equation

$$\frac{du}{dx} = F(x,u) \tag{a.0}$$

If $G$ is a one-parameter group of transformations on an open suset $M \subset X \times U \cong R^2$, let

$$v = x(x,u)\frac{\partial}{\partial x} + f(x,u)\frac{\partial}{\partial u} \tag{a.1}$$

be its infinitesimal generator. The first prolongation of $v$ is the vector field

$$pr^{(1)}v = x(x,u)\frac{\partial}{\partial x} + f(x,u)\frac{\partial}{\partial u} + f^x \frac{\partial}{\partial u_x}, \tag{a.2}$$

where

$$f^x = f_x + (f_u - x_x)u_x - x_u u_x^2. \tag{a.3}$$

Thus the infinitesimal condition that group $G$ be a symmetry group of (2.74) is

$$\frac{\partial f}{\partial x} + \left(\frac{\partial f}{\partial u} - \frac{\partial x}{\partial x}\right)F - \frac{\partial x}{\partial u}F^2 = x\frac{\partial F}{\partial x} + f\frac{\partial F}{\partial u}, \tag{a.4}$$

and any solution $x(x,u), f(x,u)$ of the partial differential equation generates a one-parameter symmetry group of our ordinary differential equation.

$$F(x,u) \equiv -\frac{(2a_1 + a_2)n}{2} \cdot x^{-2} + \frac{a_1 a_2 n^2}{2} x^{-3} \cdot u^{-1} \tag{a.5}$$

$$\frac{\partial F}{\partial x}(x,u) \equiv -(-2)\frac{(2a_1 + a_2)n}{2} \cdot x^{-3} + (-3) \cdot \frac{a_1 a_2 n^2}{2} x^{-4} \cdot u^{-1} \tag{a.6}$$

$$\frac{\partial F}{\partial u}(x,u) \equiv (-1) \cdot \frac{a_1 a_2 n^2}{2} x^{-3} \cdot u^{-2}, \tag{a.7}$$

$$F^2 = \left\{-\frac{(2a_1 + a_2)n}{2} \cdot x^{-2} + \frac{a_1 a_2 n^2}{2} \cdot x^{-3} \cdot u^{-1}\right\}^2$$
$$\equiv Ax^{-4} + Bx^{-5}u^{-1} + Cx^{-6}u^{-2} \tag{a.8}$$

where

$$A = \left[\frac{(2a_1 + a_2)n}{2}\right]^2,$$

$$B = -(2a_1 + a_2) \cdot \frac{a_1 a_2}{2} \cdot n^3,$$

$$C = \left[\frac{a_1 a_2}{2} n^2\right]^2.$$

The simplest solution could be set

$$\mathbf{x} = \mathbf{x}(x) \tag{a.9}$$

$$\mathbf{f} = \mathbf{f}(u) \tag{a.10}$$

Thus

$$\begin{aligned}
\mathbf{x} \cdot &\left\{(2a_1 + a_2)nx^3 u^2 - \frac{3}{2} a_1 a_2 n^2 x^2 u\right\} \\
+ \mathbf{f} \cdot &\left\{-\frac{1}{2} a_1 a_2 n^2 x^3\right\} \\
= (\mathbf{f}_u - \mathbf{x}_x) \cdot &\left\{-\frac{(2a_1 + a_2)}{2} nx^4 u^2 + \frac{1}{2} a_1 a_2 n^2 x^3 u\right\}.
\end{aligned} \tag{a.11}$$

These all have the general solution

$$\mathbf{x} = c_1 + c_2 x, \tag{a.12}$$

$$\mathbf{f} = c_3 + c_4 u \tag{a.13}$$

where $c_1, c_2, c_3, c_4$ are arbitrary constants.

Substituting the general formulae, we find the determining equations for the symmetry group of the scaling equation to be the following:

| Monomial | Coefficient |
|---|---|
| $x^3$ | $-\dfrac{a_1 a_2}{2} n^2 \cdot c_3 = 0$ |
| $x^2 u$ | $-\dfrac{3 a_1 a_2}{2} n^2 \cdot c_1 = 0$ |
| $x^3 u$ | $-\dfrac{3 a_1 a_2}{2} n^2 \cdot c_2 - \dfrac{a_1 a_2}{2} n^2 \cdot c_4 = \dfrac{a_1 a_2}{2} n^2 \cdot (c_4 - c_2)$ |
| $x^4 u^2$ | $(2a_1 + a_2)n \cdot c_2 + \dfrac{(2a_1 + a_2)}{2} n \cdot (c_4 - c_2) = 0$ |

The solution of the determining equations is elementary. We conclude that the most general infinitesimal symmetry of the scaling equation has coefficient functions of the form

$$\mathbf{x} = c_2 x \equiv \frac{x}{a}, \tag{a.14}$$

$$f = c_4 u \equiv \frac{u}{b}, \tag{a.15}$$

where

$$c_2 + c_4 = 0.$$

The first prolongation of $v$ is the vector field

$$pr^{(1)}v = x(x,u)\frac{\partial}{\partial x} + f(x,u)\frac{\partial}{\partial u} + f^x \frac{\partial}{\partial u_x}, \tag{a.16}$$

where

$$f^x = f_x + (f_u - x_x)u_x - x_u u_x^2.$$
$$= (c_4 - c_2)u_x$$

$$pr^{(1)}v = \frac{x}{a}\cdot\frac{\partial}{\partial x} + \frac{u}{b}\cdot\frac{\partial}{\partial u} + \frac{a-b}{ab}\cdot p\cdot\frac{\partial}{\partial p}, \tag{a.18}$$

The characteristic equation reads:

$$\frac{adx}{x} = \frac{bdu}{u} = \frac{ab}{(a-b)}\cdot\frac{dp}{p} \tag{a.19}$$

One can obtain the invariants as follows:

$$I_1 = \frac{u^b}{x^a}, \tag{a.20}$$

$$I_2 = \frac{p}{x^{\frac{a}{b}-1}}. \tag{a.21}$$

It should be noted that

$$u = [I_1]^{\frac{1}{b}} x^{\frac{a}{b}} \tag{a.22}$$

$$p = \frac{du}{dx} = I_2 x^{\frac{a}{b}-1} \tag{a.23}$$

We also know that

$$c_2 + c_4 = \frac{1}{a} + \frac{1}{b} = 0. \tag{a.24}$$

$$\frac{b}{a} = -1. \tag{a.25}$$

.

Hence, we have

$$P(l) = \frac{dl}{dt} = [I_1]^{\frac{1}{b}}\cdot l^{-1} \tag{a.26}$$

$$ldl = [I_1]^{\frac{1}{b}}\cdot dt \tag{a.27}$$

$$l^2 = [I_1]^{\frac{1}{b}} \cdot (t + t_0). \tag{a.28}$$

# Appendix 2. Solutions of the Lienard equation

For the general Lienard type equation,
$$y''_{xx} + f(y)y'_x + g(y) = 0. \tag{a.2.1}$$

one can choose the below variable transformation
$$h(y) = y'_x \tag{a.2.2}$$

$$y''_{xx} = \frac{dh}{dx} = \frac{dh}{dy} \cdot \frac{dy}{dx} = hh'_y \tag{a.2.3}$$

The substitution of this leads to a first order equation:
$$hh'_y = -f(y)h - g(y) \tag{a.2.4}$$

This equation belongs to the secondary kind (special case) Abel equation.
The substitution
$$z = -\int f(y)dy \tag{a.2.5}$$

brings the Abel equation to the canonical form:
$$hh'_z = h + \Phi(z) \tag{a.2.6}$$

where the function $\Phi(z)$ is defined parametrically by the relations

$$\Phi(z) = \frac{g(y)}{f(y)} \tag{a.2.7}$$

Solvable Abel equations of this form can be obtained by using the results presented in the books by Zaitsev and Polyanin (1994) and Polyanin & Zaitsev (2003). They presented a large number of exact solutions to the Abel equation of the secondary kind of various $f(x), g(x)$.

Consider the Abel equation
$$hh'_z = a_1 y^{n-1} h + a_2 y^{2n-1} \tag{a.2.8}$$

where
$$a_1 = a(2n+k)y^k + b \tag{a.2.9}$$

$$a_2 = -a^2 nx^{2k} - abx^k + c \tag{a.2.10}$$

The substitution

$$h = y^n \left(w + ay^k\right) \tag{a.2.11}$$

leads to a Bernoulli equation with respect to $y = y(w)$:

$$\left(nw^2 - bw - c\right)y'_w = -wy - ay^{k+1}. \tag{a.2.12}$$

[1] Zaitsev, V.F. and Polyanin, A.D., Discrete-Group Methods for Integrating Equations of Nonlinear Mechanics, CRC Press, Boca Raton, 1994.
[2] Polyanin, A.D. and Zaitsev, V.F., Handbook of Exact Solutions for Ordinary Differential Equations. 2nd Edition, Chapman & Hall/CRC, Boca Raton, 2003.